\documentclass[preprint2]{aastex6}

\def\4c50{4C 50.11}
\def\fermi{{\it Fermi}}

\begin{document}


\title{High energy properties of the flat spectrum radio quasar 4C~50.11}


\author{Jia-Neng Zhou\altaffilmark{1}, V. Jithesh\altaffilmark{2,1}, Liang Chen\altaffilmark{1,3}, Zhongxiang Wang\altaffilmark{1}}


\altaffiltext{1}{Shanghai Astronomical Observatory, Chinese Academy of Sciences, 80 Nandan Road, Shanghai 200030, China;zjn@shao.ac.cn}
\altaffiltext{2}{Inter-University Centre for Astronomy and Astrophysics, Post Bag 4, Ganeshkhind, Pune 411007, India}
\altaffiltext{3}{University of Chinese Academy of Science,  19A Yuquanlu, Beijing 100049, China}

\begin{abstract}

We investigate the $\gamma$-ray and X-ray properties of the Flat Spectrum
Radio Quasar (FSRQ) \4c50 at redshift $z= 1.517$. The {\it Fermi}-LAT data
indicate that this source was in an active state since
2013 July. During this active period,
the source's emission appeared harder in $\gamma$-rays, with
the flux having increased by more than a factor of three.
We analyze two distinct flares seen in the active state
and find that the variability is as short as several hours.
The {\it Swift}-XRT data show that the source was
variable at X-ray energies, but no evidence was found for flux or
spectral changes
related to the $\gamma$-ray activity. The broad-band X-ray spectrum obtained
with {\it Swift}-XRT and {\it NuSTAR} is well described by a broken PL model,
with an extremely flat spectrum ($\Gamma_{1} \sim 0.1$) below the break energy,
$E_{\rm break} \sim 2.1~{\rm keV}$, and $\Gamma_{2} \sim 1.5$ above
the break energy. The spectral flattening below $\sim 3$ keV is likely due
to the low energy cut-off in the energy distribution of the photon-emitting
electron population.
We fit the broad-band spectral energy distribution of the source during
both the active and quiescent states.
The X-ray and $\gamma$-ray emission from the jet is mainly due to
the inverse-Compton scattering process, with the seed photons provided
from the broad line region, and the jet is estimated to be larger than
the accretion power if the jet is mainly composed of electron-proton pairs.
\end{abstract}




\section{Introduction} \label{sec:intro}

Blazars are radio-loud active galactic nuclei (AGNs) with relativistic
jets pointing towards the Earth \citep{1978bllo.conf..328B}.
Because of the Doppler beaming effect, emission from a jet dominates
the broad-band spectral energy distribution (SED) from radio to
$\gamma$-rays energies \citep{1995PASP..107..803U}. The SEDs usually have two broad bumps
in a $\log\nu-\log\nu f_{\nu}$ diagram. While the low-energy bump usually peaks
from infrared to X-ray energies, which are believed to be the synchrotron
emission of non-thermal electrons, the high-energy bump peaks from
X-ray to $\gamma$-ray bands, which is considered to be the inverse
Compton (IC) emission
of the same electron population. For the IC emission, the seed photons can
come from the low-energy synchrotron emission, broad line region (BLR), or
dusty torus \citep[see e.g.,][]{1981ApJ...243..700K, 1985ApJ...298..128B, 1992ApJ...397L...5M, 2000ApJ...545..107B}. Because of the synchrotron self-absorption effect, blazars
tend to have flat radio spectra with spectral index $\alpha < 0.5$.
As a subclass of blazars, flat spectrum radio quasars (FSRQs) have strong
optical emission lines (equivalent width $>5${\AA}), comparing to
BL Lac objects that show no or very weak emission lines \citep{1997A&A...325..109S}.

In the current third \fermi\ Large Area Telescope (LAT) source catalogue (3FGL),
the dominant extragalactic $\gamma$-ray sources are
blazars \citep{2015ApJS..218...23A}. Extreme variability is not common
to all blazars detected in $\gamma$-rays.
The minimal variable timescale detected
with {\it Fermi}-LAT has reached less than half an hour
(e.g. PKS 1510-089, \citealp{2013A&A...555A.138F}) and the variation
amplitude can be two orders of magnitude
(e.g., 3C 454.3; \citealp{2011ApJ...733L..26A}). Detailed studies of
spectra and variabilities are essential for determining the location
and mechanism of radiation from the jets of the blazars.

The FSRQ \4c50 (also known as NRAO 150) is one of the strongest radio and
millimeter AGN sources in the northern sky
\citep{1966ApJS...13...65P,2008ASPC..386..249A, 2010ApJS..189....1A}.
The VLBI monitoring observations showed that the inner jet
(inner 0.5 mas from the core) exhibits superluminal motions with
$\beta_{app} \sim (6.3\pm1.1)c$ and a large, $>100^{\circ}$ projected misalignment
of the jet within the inner 0.5 mas to 1 mas from the
core \citep{2007A&A...476L..17A, 2014A&A...566A..26M}. These properties
imply that a relativistic jet points toward the Earth with a very small
viewing angle \citep{2007A&A...476L..17A}.
\citet{2010A&A...519A...5A} measured the redshift using near-IR spectroscopic
data (exhibiting strong H$\alpha$ and H$\beta$ emission lines), and derived
the cosmological redshift $z=1.517 \pm 0.002$, which corresponds to the
luminosity distance $d_{L}=11.2 \times 10^{3}~{\rm Mpc}$.
\citet{2010ATel.2517....1F} reported
the detection of $\gamma$-ray emission from \4c50 with LAT on board the
{\it Fermi} satellite. Using almost 20 months
of data, he provided the $\gamma$-ray flux above 100 MeV,
$F_{100 MeV} = 3.2 \pm 1.1 \times 10^{-8}~{\rm photons}~{\rm cm}^{-2}~{\rm s}^{-1}$,
and photon index $\Gamma = 2.6 \pm 0.2$. After the $\gamma$-ray flaring
activity around $\sim$MJD 56686 (2014 January 29),
{\it Swift} target-of-opportunity
observations were performed \citep{2014ATel.5838....1C, 2014ATel.5878....1K}.

For the purpose of fully studying this high-energy source,
we collected its \fermi-LAT and available X-ray data, which include
15 {\it Swift} observations and one {\it NuSTAR} observation,
and performed detailed analysis of the data.
In this paper, we present the results from our analysis.
In the following, \S~2 describes the
data analysis of the {\it Fermi}-LAT, {\it Swift} and {\it NuSTAR} observations. 
The obtained temporal and spectral results
are presented in \S~3 and \S~4, respectively. We discuss the overall properties of the
source in \S~5, including fitting to its multiwavelength SED, and we summarize our results in
\S~6.

\section{Data Reduction}

\subsection{{\it Fermi}-LAT data analysis}

We used approximately seven-years {\it Fermi}-LAT Pass 8 data in this work,
which are
from MJD 54682 (2008-08-04) to MJD 57352 (2015-11-26),  with the energy range
from 100 MeV to 100 GeV. During the time period, \4c50 was in an active
state from MJD 56482 (2013-07-09) to the end of the data. The Fermi Science
Tools v10r0p5 package was used to analyse the data, with
the {\tt P8R2\_SOURCE\_V6} instrument response functions (IRFs) applied.
To avoid
contamination from $\gamma$-rays reflected by the Earth, we selected the
events with zenith angles $\le 90^{\circ}$.

In the analysis, photons from a 20$^{\circ}\times 20^{\circ}$ square
region of interest (ROI) centred at the position of \4c50 were selected,
and binned into spatial pixels of 0.1$^{\circ}\times 0.1^{\circ}$.
The first run of the analysis, using {\tt gtlike},
was performed with the binned likelihood method
to derive the sky map model. We modelled the events considering
the components of the target and background. The background was composed
of sources in the 3FGL catalogue \citep{2015ApJS..218...23A} within the ROI and
diffuse components. The latter included the Galactic diffuse
model ({\tt gll\_iem\_v06.fits}) and isotropic
background ({\tt iso\_P8R2\_SOURCE\_V6\_v06.txt}).

To confirm the spatial association of $\gamma$-ray emission with \4c50,
we calculated a $2^{\circ}\times 2^{\circ}$ Test Statistic (TS) map
centred at its coordinates.
A putative point source was assumed and moved through a grid
of locations on the sky by maximizing $-\log$(likelihood) at each grid point.
In this step, the target source that corresponds to \4c50 was unmodelled
(i.e., removed from the model file). All parameters of point sources in
the ROI, except the diffuse components, were fixed at the 3FGL catalogue
values.
To reduce the contamination due to the large point spread function (PSF) at low
energies, we only used photons above 1 GeV for TS map calculation.
The $\gamma$-ray source was significantly detected, with a maximum TS value
of $\simeq 283$.
We derived the position of the source, and the best-fit position is
RA = $59.872^{\circ}$, Decl. = $50.968^{\circ}$ (J2000.0) with a
positional uncertainty of $0.022^{\circ}$.
The derived position is only 0.25$'$ away from \4c50.

Since \4c50 is located on the Galactic plane (\textit{b}=$-1.6^{\circ}$;
\citealt{2004AJ....127.3587F}), we checked the SIMBAD database for sources
within the error circle.
There are only a few sources catalogued by 2MASS and SDSS (a ROSAT source
1RXS J035930.6+505730 is also within the error circle, but it has been
considered as the likely counterpart to \4c50; \citealt{2007A&A...476L..17A}).
The nearest source is located at a distance of $3.6^{\circ}$ \citep{2015ApJS..218...23A}.
In addition, since the $\gamma$-ray background can be
complex at the Galactic plane, we also checked the normalization
value for the Galactic diffuse emission. It was 0.963$\pm$0.003.
Considering the systematic uncertainty of
6\% for the background \citep{2013ApJS..208...17A,2010ApJ...722.1303A},
the value is consistent with the expectation of the normalization of 1.

\subsection{{\it Swift} Data Analysis}

The {\it Swift} satellite \citep{2004ApJ...611.1005G} performed 15
observations of \4c50 between 2007 January and 2015 December. We utilized
archival data from the X-ray telescope (XRT; \citealt{2005SSRv..120..165B}) on board {\it Swift}.
The XRT data were processed with standard filtering and screening criteria,
using the {\sc xrtpipeline} version 0.13.0 in the {\sc heasoft} package
version 6.15.1. The photon-counting (PC) mode data were collected from
all the observations. Since the source had low count rates
($< 0.1~\rm counts~s^{-1}$), the pile-up correction was not required.
Source events were extracted from a circular region with a radius
of 47\arcsec, while background events were extracted from a circular region
of the same radius, with the standard grade filtering of 0--12.
We generated the ancillary response files with the tool {\sc xrtmkarf}
and used the spectral redistribution matrices available in the calibration
data base (CALDB) version 20151105. The spectra were binned to contain
at least 20 counts per bin, which allowed the $\chi^2$ spectral fitting.
In other cases, where there were no sufficient spectral counts,
the Cash Statistics \citep{1979ApJ...228..939C} was used for spectral
modelling.

\subsection{{\it NuSTAR} Data Analysis}

{\it NuSTAR} \citep{2013ApJ...770..103H} observed \4c50 with
its focal plane module A (FPMA) and B (FPMB) X-ray telescopes,
on 2015 December 14 for an exposure time of 20.5 ks
(Observation ID: 60160177002). We processed the data
with the {\it NuSTAR} Data Analysis Software {\sc nustardas} version 1.3.1.
We cleaned and calibrated the unfiltered event files using standard
filtering criteria with the {\sc nupipeline} task and {\it NuSTAR} CALDB
version 20151008. The source and background regions were taken
from a circular region of radius $70 \arcsec$, and we generated the spectra,
response matrices, and ancillary response files,
using {\sc nuproducts} for both focal plane modules (FPMA and FPMB).
The {\it NuSTAR} FPMA and FPMB spectra were grouped with a minimum of
20 counts per bin using {\sc heasoft} task {\sc grppha}. We did not
combine the spectra from FPMA and FPMB; instead we jointly fit
the two spectra.

\section{Variability in $\gamma$-ray and X-ray}

\subsection{$\gamma$-ray temporal properties}

Since \4c50 was active from $\sim$MJD 56482, showing $\gamma$-ray flares,
a 5-days binned light curve was derived using the binned likelihood method.
Normalizations of all point sources within 5$^{\circ}$ from the target
and sources with variable
index\footnote{\url{http://fermi.gsfc.nasa.gov/ssc/data/access/lat/4yr_catalog/}} $\ge$ 72.44
were set free. For the purpose of studying the flaring variability in detail,
24-, 12-, and 6-hours binned light curves were also created. For these
light curves, an unbinned likelihood method was used due to the low statistics,
and only the normalizations of variable sources were set free
in the background model.
In the analysis, when a data point had TS$<$5, we calculated its flux
upper limit at a 95\% confidence level \citep{2011ApJ...736L..11A}.

\begin{figure}
\begin{center}
{\includegraphics[width=0.8\linewidth]{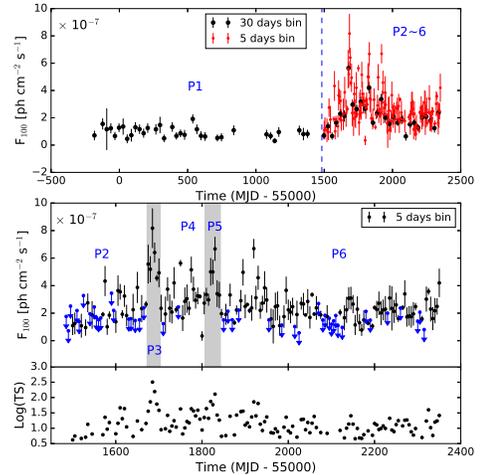}}
\end{center}
\caption{\fermi-LAT light curves of \4c50 with energy between 100 MeV and
100 GeV. Different time intervals are shown
based on the light curve properties. {\it Upper:} Long-term light
curves binned with 30 days (black points) and 5 days (red points).
{\it Lower:} 5-day binned light curve of \4c50 during the $\gamma$-ray
active period
from MJD 56482, and TS values for each bin. When TS $<$ 5, only flux
upper limits (95\% confidence) are obtained.}
\label{lc_phase}
\end{figure}

\begin{figure}
\begin{center}
{\includegraphics[width=0.8\linewidth]{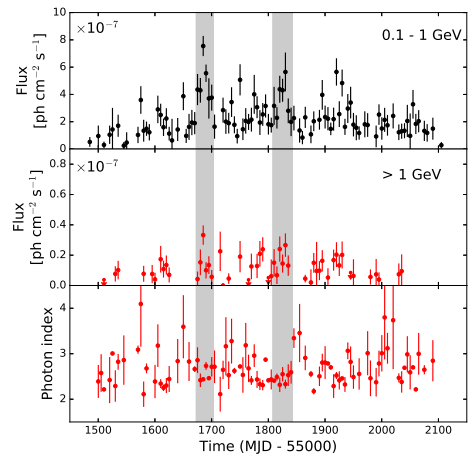}}
\end{center}
\caption{5-day binned light curves of \4c50\ observed by {\it Fermi}-LAT during
the $\gamma$-ray active period. From top to bottom:
light curve at 0.1--1 GeV band, 1--100 GeV band,
and photon index obtained from a power-law model.
The shaded area indicates $\gamma$-ray flares P3 and P5.}
\label{dif_lc}
\end{figure}

Figure~\ref{lc_phase} shows the $\gamma$-ray light curves of \4c50
in 0.1--100 GeV, with the upper panel covering the
entire {\it Fermi} observation time period (setting 30 day time bins)
and the lower panel covering the $\gamma$-ray active period starting
from $\sim$MJD 54682.
Six time intervals are shown based on the light
curves (see Figure~\ref{lc_phase}): P1 is the time period before
the active state and P2--P6
the time periods covering the active state, which lasts for more than two years.
Examining the light curves, there are two distinct $\gamma$-ray flares in
the active state, P3 and P5, which appear to contain several data points
above the nearby flux levels and have peak fluxes more than two times higher.
We thus obtained smoothed 5-day light curves (by shifting each time bin
by 1 day) for them and determined their approximate time durations
(marked by the grey area in Figure~\ref{dif_lc}).
Detailed analysis of the peak regions of the two flares are provided below.

We investigated the active state of \4c50 in different energy bands.
The two flaring events (P3 and P5) are clearly seen in the 0.1--1 GeV energy
band (Figure~\ref{dif_lc}).
However in the $> 1$ GeV energy light curve, the flares do not have
significantly higher flux than the rest of the light curve,
and the overall photon flux is an order of
magnitude lower than that in the 0.1--1 GeV energy band.

We modelled the photons in each 5-day time bin with a single power
law (PL) model because of the low counts of the data points
(see below for detailed discussion about spectral
modelling in Section~\ref{subsec:sa}).
As shown in the bottom panel of
Figure~\ref{dif_lc}, the derived photon index varies with time and flux.
However no obvious trend can be drawn from the variations, as
the photon index generally has values in a range of 2--3.

\begin{figure*}
\begin{center}
{\includegraphics[width=0.4\linewidth]{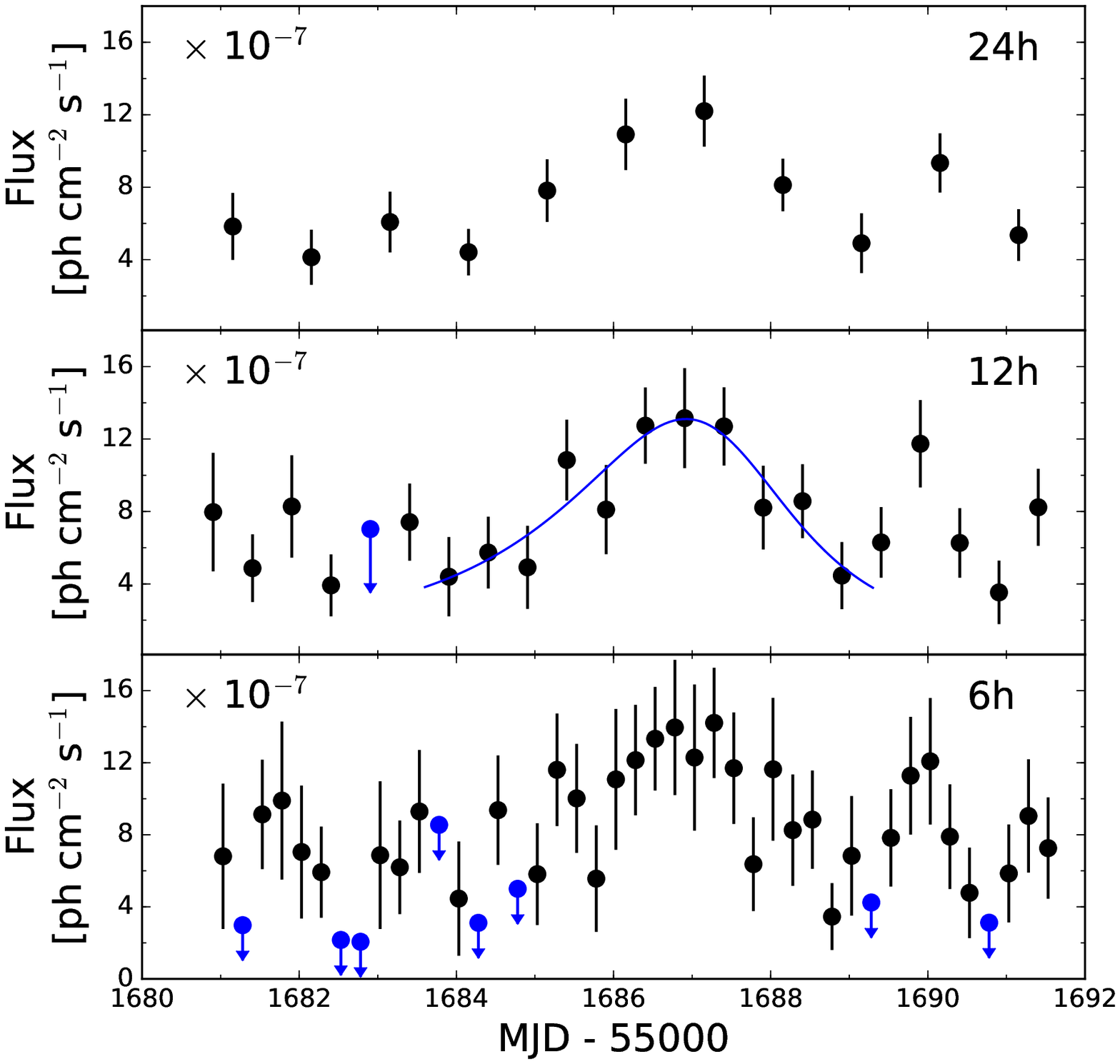}
\includegraphics[width=0.4\linewidth]{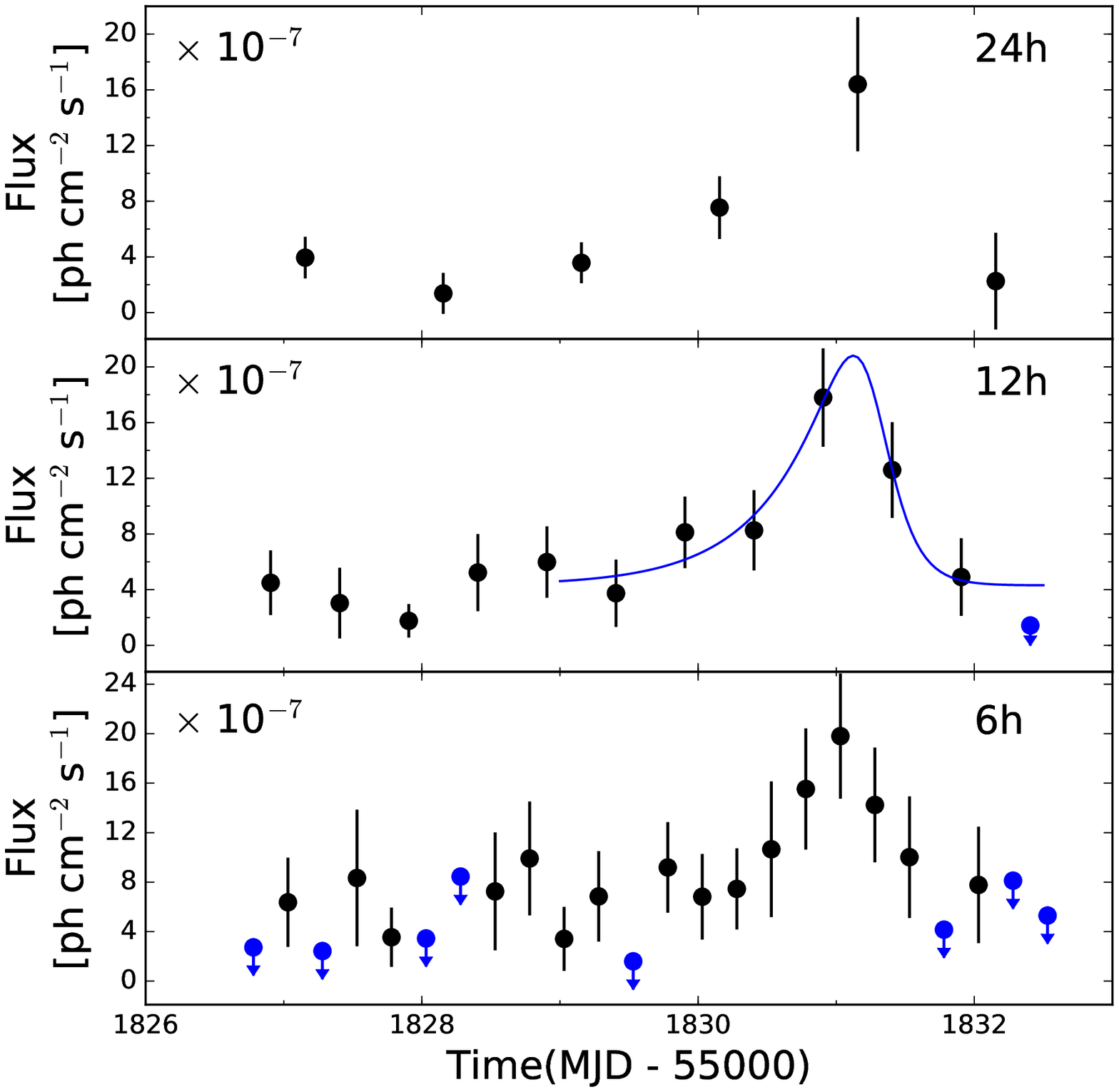}}
\end{center}
\caption{Light curves of the peak regions in flares P3 ({\it left}) and
P5 ({\it right}).
From {\it top} to {\it bottom}: the light curves binned with 24,
12, and 6 hours, respectively. For points with TS $<$ 5, upper limits are
calculated at a 95\% confidence level. The curves in the middle panels
are the fits to the flux variations.}
\label{lc_flares}
\end{figure*}

The peak regions of the two flares are shown in Figure~\ref{lc_flares}.
The first
one started at $\sim$ MJD 56685 and lasted for $\sim$5 days. During this
period, the profile is relatively flat, while a small sub-flare
around MJD 56690 is seen, which lasted only $\sim$2 days. We extracted
the spectrum of the peak region and fitted it with a single PL model. From
fitting, we found averaged photon flux
$F_{100 MeV} = 6.3\pm0.3 \times 10^{-7}~{\rm photons}~{\rm cm}^{-2}~{\rm s}^{-1}$
and photon index $\Gamma=2.55\pm0.04$. In the second flare, the peak is
around $\sim$MJD 56831, which has a different shape compared to that of
the first one. The averaged photon flux of this flare was
$F_{100 MeV} = 4.6\pm0.3 \times 10^{-7}~{\rm photons}~{\rm cm}^{-2}~{\rm s}^{-1}$
and photon index $\Gamma=2.44\pm0.05$. The peak flux (given from the 6-hours light curves) was
$2.0\pm0.5 \times 10^{-6}~{\rm photons}~{\rm cm}^{-2}~{\rm s}^{-1}$. We
studied the two peaks by fitting their light curves with Equation 7
in \citet{2010ApJ...722..520A},

\begin{equation}
F(t)=F_{c}+F_{0}[e^{({t_{0}-t)}/\tau_{r}}+e^{(t-{t_{0})}/\tau_{d}}]^{-1}\ \ \ ,
\label{lcfiteq}
\end{equation}
which is widely used to characterize a variation
profile \citep[see e.g.][]{2015ApJ...807...79H}. In this function,
$F_{c}$ and $F_{0}$ are the underlying constant level and flare amplitude,
respectively, $t_{0}$ approximately corresponds to the flux peak time (when
a flare has a symmetric shape), and raise time $\tau_{r}$ and
decay time $\tau_{d}$ characterize the time scales for the raising
and decaying parts of a flare. We chose to fit the peaks in the middle
panels of Figure~\ref{lc_flares}, which
are significant and are relatively well resolved.
The fitting results are listed in
Table~\ref{flare_fit}. The two peaks in P3 and P5 show asymmetric
profiles, with a minimum time scale in P5 as short as $\sim$4\ hr.

\begin{table*}
\centering
\caption{Fitting results for the peaks in flares P3 and P5.}
\begin{tabular}{@{}lcccrr@{}}
\hline
\hline
Flare  & $F_{c}$ & $F_{0}$ & $t_{0}$ & $\tau_{r}$ & $\tau_{d}$  \\
  & ($\times10^{-7}$) & ($\times10^{-7}$)  & (MJD) &    (hr)  &  (hr) \\
\hline
P3 & 1.3$\pm$0.3 & 22.6$\pm$0.4 & 56687.27$\pm$0.73 & 40.1$\pm$1.4 & 22.3$\pm$0.5 \\
P5 & 4.3$\pm$0.3 & 28.2$\pm$1.4 & 56831.26$\pm$0.04 & 11.9$\pm$0.9 & 3.5$\pm$1.1 \\
\hline
\end{tabular}
\label{flare_fit}
\tablecomments{$F_{c}$ and $F_{0}$ are in units of ${\rm ph}~{\rm cm}^2~{\rm s}^{-1}$.}
\end{table*}

\subsection{X-ray temporal properties}

\begin{figure}
\begin{center}
{\includegraphics[width=0.8\linewidth,angle=0]{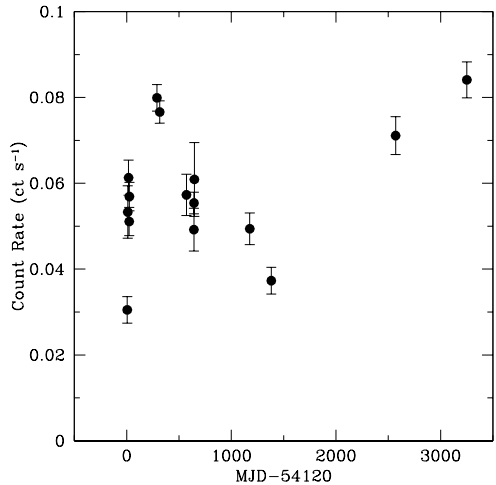}}
\end{center}
\caption{Count rate variations of \4c50 seen in the
{\it Swift}-XRT observations over year 2007--2015.}
\label{lc}
\end{figure}

The obtained long-term {\it Swift}-XRT count-rate curve of \4c50 is shown in
Figure \ref{lc}.  During year 2007--2015, the source showed variability
with the count rates varying by a factor of $\sim 3$. We also observed an
increase of the count rate in the latest {\it Swift}-XRT observations
and it reached the maximum value of
$\sim 0.08~\rm counts~s^{-1}$ on 2015 December 14.
Since the $\gamma$-ray analysis points to hour-scale variability of the
source, we checked the longest {\it Swift}-XRT observation
(conducted on 2007 December 02), but did not find any evidence for such
variability in light curves binned at different half-hourly or hourly
timescales.

\section{Spectral Properties}

\subsection{Spectral Properties in $\gamma$-ray band}
\label{subsec:sa}

\begin{figure}
\begin{center}
{\includegraphics[width=0.95\linewidth]{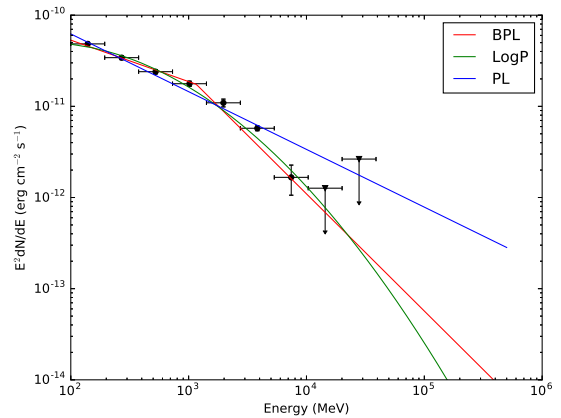}}
\end{center}
\caption{\fermi\ $\gamma$-ray spectrum of \4c50 during the total P2--P6 time interval. The PL, BPL, and LogP model fits are shown as blue, red, and green curves, respectively.}
\label{model_comparison}
\end{figure}

We analysed the $\gamma$-ray spectra of \4c50 in the different time
intervals (P1--P6) and the total time interval of the combined P2--P6.
Models of simple PL ($dN/dE \propto E^{-\Gamma}$),
broken PL (BPL; $dN/dE \propto E^{-{\Gamma}_{1}}$ for $E<E_{break}$,
and $dN/dE \propto E^{-{\Gamma}_{2}}$ for $E>E_{break}$),
and log-parabola
($dN/dE\propto\left(E/E_{b}\right)^{-\alpha-\beta\log\left(E/E_{b}\right)}$)
were considered. The results are given in Table~\ref{gamma}.
The three models generally describe the spectrum well, as indicated by the obtained TS values that are nearly the same. However for the total time interval of P2--P6, the active state, the BPL and log-parabola
models are probably more favoured than the single PL.
In Figure~\ref{model_comparison}, we showed the spectrum from the combined
P2--P6 data and the three model fits. The PL fit does not describe the high
energy tail of the spectrum as well as the two other models, which is
also supported by values of likelihood ratio $-2{\Delta}L$
(see Table~\ref{gamma}).

From the analysis, one property may be drawn if we consider the results
from the PL fits or BPL fits: emssion from the source appears harder
when brighter (see Table~\ref{gamma}). However, the uncertainties are too
large, not allowing us to have a clear conclusion. In any case, it is certain
that emission in the active state is harder than that in quiescence.

\begin{table*}
\small
\begin{center}
\caption{$\gamma$-ray spectral fitting results for \4c50 in different time intervals.}
\begin{tabular}{@{}clllllll@{}}
\hline
\hline
 Interval & Model & $\alpha/\Gamma/\Gamma_1$ & $\beta/\Gamma_2$ & E$_{break}$ & TS & $-2{\Delta}L$ & Flux \\
  &  &  &  & (GeV) & & & ($10^{-7}$)\\
\hline
 & PL   & 3.01$\pm$0.03 & ...     & ...   & 303 & ... & (6.99$\pm$0.35)E-1 \\
 P1 & LogP & 3.01$\pm$0.08 & 0.00$\pm$0.00 & ...   & 304 & 0.0 & (6.99$\pm$0.48)E-1 \\
  & BPL  & 2.97$\pm$0.17& 3.15$\pm$0.77& 1.0$\pm$0.1& 303 & 0.2 & (6.94$\pm$0.45)E-1 \\\hline
  & PL   & 2.63$\pm$0.07 & ...     & ...   & 234 & ... & 1.65$\pm$0.15 \\
 P2 & LogP & 2.23$\pm$0.13 & 0.18$\pm$0.06 & ...   & 236 & 6.4 & 1.50$\pm$0.15 \\
  & BPL  & 2.38$\pm$0.11& 3.30$\pm$0.43& 1.0$\pm$0.3& 237 & 6.8 & 1.52$\pm$0.16 \\\hline
  & PL   & 2.55$\pm$0.04 & ...     & ...   & 654 & ... & 6.30$\pm$0.30 \\
 P3  & LogP & 2.42$\pm$0.10 & 0.06$\pm$0.04 & ...   & 651 & 3.8 & 6.14$\pm$0.39 \\
  & BPL  & 2.44$\pm$0.08& 2.87$\pm$0.24& 1.0$\pm$0.2& 652 & 5.0 & 6.11$\pm$0.39 \\\hline
  & PL   & 2.61$\pm$0.06 & ...     & ...   & 347 & ... & 2.87$\pm$0.21 \\
 P4  & LogP & 2.17$\pm$0.13 & 0.22$\pm$0.06 & ...   & 304 & 10.6 & 2.63$\pm$0.21 \\
  & BPL  & 2.37$\pm$0.09& 3.40$\pm$0.47& 1.0$\pm$0.3& 303 & 9.2 & 2.69$\pm$0.22 \\\hline
  & PL   & 2.44$\pm$0.05 & ...     & ...   & 358 & ... & 4.59$\pm$0.28 \\
 P5  & LogP & 2.22$\pm$0.08 & 0.10$\pm$0.03 & ...   & 359 & 2.4 & 4.37$\pm$0.29 \\
  & BPL  & 2.26$\pm$0.07& 2.90$\pm$0.22 & 1.0$\pm$0.0& 361 & 3.8 & 4.35$\pm$0.34 \\\hline
 & PL   & 2.69$\pm$0.05 & ...     & ...   & 504 & ... & 2.10$\pm$0.12 \\
 P6 & LogP & 2.51$\pm$0.09 & 0.08$\pm$0.04 & ...   & 503 & 3.2 & 2.03$\pm$0.13 \\
  & BPL  & 2.58$\pm$0.07& 2.97$\pm$0.21& 1.0$\pm$0.2& 504 & 2.6 & 2.05$\pm$0.13 \\\hline
  & PL   & 2.63$\pm$0.02 & ...     & ...   & 2175 & ... & 2.38$\pm$0.07 \\
 P2--P6 & LogP & 2.35$\pm$0.05 & 0.14$\pm$0.02 & ... & 2180 & 32.2 & 2.24$\pm$0.07 \\
  & BPL & 2.45$\pm$0.04 & 3.29$\pm$0.17 & 1.2$\pm$0.1 & 2191 & 36.4 & 2.26$\pm$0.07 \\
\hline
\end{tabular}
\label{gamma}
\end{center}
\tablecomments{
Flux (0.1--100 GeV) is in units of
$~{\rm ph}~{\rm cm}^{-2}~{\rm s}^{-1}$. $-2{\Delta}L$ is the difference of
log(Likelihood) of the model with respect to that of the single power law model
(e.g., \citealt{2010ApJ...721.1425A}).}
\end{table*}

\subsection{Spectral Properties in X-ray band}
\label{subsec:joint}

\begin{table*}
\small
\begin{center}
\caption{{\it Swift}-XRT observations of \4c50 and
the fitting results using a PL model with $N_{\rm H}$ fixed to the Galactic absorption}
\begin{tabular}{@{}lccccr@{}}
\hline
\hline
ObsID & Date & Exposure time & $\Gamma_{\rm X}$ & $0.3-10$ keV Flux & $\chi^2/d.o.f$ \\
      &      & (s)           &          & ($\rm 10^{-12}\ erg~cm^{-2}~s^{-1}$) & \\
\hline
00030879001 & 2007 Jan 25 & 3505  & $0.76^{+0.44}_{-0.44}$ & $5.36^{+1.75}_{-1.23}$ & 7.6/7(C) \\
00030879002 & 2007 Jan 30 & 1513  & $1.46^{+0.47}_{-0.47}$ & $6.65^{+1.42}_{-1.24}$ & 5.6/12(C) \\
00030879003 & 2007 Feb 07 & 3845  & $1.00^{+0.28}_{-0.28}$ & $8.06^{+1.27}_{-1.14}$ & 3.0/8 \\
00030879004 & 2007 Feb 13 & 5329  & $1.17^{+0.27}_{-0.27}$ & $6.38^{+0.86}_{-0.81}$ & 3.2/8 \\
00030879005 & 2007 Feb 15 & 5276  & $1.16^{+0.24}_{-0.24}$ & $6.48^{+0.78}_{-0.74}$ & 9.0/11 \\
00036308001 & 2007 Nov 05 & 8778  & $1.20^{+0.13}_{-0.13}$ & $8.47^{+0.63}_{-0.62}$ & 34.0/30 \\
00036308002 & 2007 Dec 02 & 11952 & $1.05^{+0.11}_{-0.11}$ & $9.02^{+0.61}_{-0.59}$ & 49.6/40 \\
00036308003 & 2008 Aug 13 & 2612  & $0.83^{+0.35}_{-0.36}$ & $7.99^{+1.94}_{-1.43}$ & 11.1/11(C) \\
00036308004 & 2008 Oct 23 & 2173  & $1.17^{+0.52}_{-0.51}$ & $7.02^{+2.05}_{-1.37}$ & 14.4/7(C) \\
00036308005 & 2008 Oct 24 & 9514  & $1.10^{+0.14}_{-0.14}$ & $7.10^{+0.61}_{-0.59}$ & 18.7/23 \\
00036308006 & 2008 Oct 27 & 873   & $0.80^{+0.71}_{-0.71}$ & $8.79^{+3.59}_{-2.30}$ & 3.3/7(C) \\
00030879006 & 2010 Apr 10 & 3780  & $0.82^{+0.29}_{-0.29}$ & $6.70^{+1.40}_{-1.09}$ & 6.1/15(C) \\
00030879007 & 2010 Nov 01 & 3930  & $0.96^{+0.36}_{-0.36}$ & $6.06^{+1.36}_{-1.01}$ & 8.5/11(C) \\
00036308007 & 2014 Feb 02 & 3854  & $1.14^{+0.23}_{-0.23}$ & $8.27^{+1.04}_{-0.99}$ & 9.3/10 \\
00080948001 & 2015 Dec 14 & 5011  & $1.25^{+0.17}_{-0.17}$ & $9.08^{+0.87}_{-0.85}$ & 27.5/17 \\
\hline
\end{tabular}
\label{xray}
\end{center}
\tablecomments{
(1) Observation ID used for the analysis; (2) date of observation; (3) exposure time in seconds for each observation; (4) photon index; (5) unabsorbed flux in $0.3-10$ keV band derived using {\tt cflux} model; (6) the $\chi^2/\rm d.o.f$ value for the model, where C-statistics is indicated by C.}
\end{table*}

We fitted 15 {\it Swift}-XRT spectra with an absorbed
PL model in the $0.3-10$ keV energy band. The absorption was incorporated
by using the photoelectric absorption model
{\tt tbabs} \citep{2000ApJ...542..914W}, which was fixed at the
Galactic value, $6.93 \times 10^{21} \rm cm^{-2}$ \citep{2005A&A...440..775K}.
The results are reported in Table \ref{xray}. All errors are given at
the 90\% confidence level.

For four observations, conducted on 2007 November 5 (Obs ID: 00036308001),
2007 December 2 (Obs ID: 00036308002), 2008 October 24 (Obs ID: 00036308005),
and 2015 December 14 (Obs ID: 00080948001), there are enough spectral counts
to test other spectral models. We initially added an extra absorption
component at the redshift of the source ({\tt ztbabs}) to the PL model.
This combined model improved the spectral fit for three observations
(with $\Delta\chi^2 \sim 7-9$) for the loss of one degree of freedom
(d.o.f) over
the single PL. For the remaining one (2008 October 24), the spectral fit
was marginally improved, $\Delta\chi^2 \sim 2$. In order to determine
the significance of the added extra absorption component,
we simulated 1000 spectra using the Monte Carlo method with
the {\sc xspec} tool {\it simftest} and fit them with the
absorbed PL and PL plus extra absorption models. An analysis of F-test
probability using this method suggests that the significance of extra
component is $2.8-3.3\sigma$ in the three observations, while the
significance is lower ($\sim 1.6\sigma$) in the observation conducted
on 2008 October 24.
The extra absorption at the redshift of the source ($\rm N_{\rm H}^{\it z}$)
is in a range of $6.5 - 12.6 \times 10^{22}~\rm cm^{-2}$ for these
four observations.

We also considered a broken PL model for all the four observations.
For the 2007 November observation, the broken PL fit resulted in
$\Gamma_{1} < 0.99$ below the break energy
$E_{\rm break} = 2.56^{+0.58}_{-1.23}$ keV and
$\Gamma_{2} = 1.68^{+0.36}_{-0.42}$ above the break energy,
($\chi^2/\rm d.o.f$ = 24.8/28). For the 2007 December observation,
the broken PL provided an acceptable fit with
$\Gamma_{1} = 0.12^{+0.58}_{-0.70}$, $E_{\rm break} = 2.08^{+0.61}_{-0.24}$ keV,
and $\Gamma_{2} = 1.35^{+0.20}_{-0.19}$ ($\chi^2/\rm d.o.f$ = 38.9/38).
The photon index $\Gamma_{1} = -0.31^{+1.10}_{-1.20}$,
$E_{\rm break} = 2.18^{+0.91}_{-0.25}$ keV, and
$\Gamma_{2} = 1.80^{+0.59}_{-0.33}$ were obtained for the observation
in 2015 ($\chi^2/\rm d.o.f$ = 15.8/15). This model provided an
improvement to the spectral fit for the three observations,
with $\Delta\chi^2 \sim 9-11$ for the loss of two extra degrees of freedom
at a probability of $> 98\%$ (from F-test), over the single PL fit.
The F-test results suggest that the broken PL is the best-fit spectral model
for the source. However for the 2008 October observation, the broken PL
model provided a marginally improved spectral fit
(at a probability of $< 80\%$) over the single PL. Moreover,
the break energy obtained ($E_{\rm break} \sim 3.26$ keV) in this observation
was not well constrained. We thus fixed $E_{\rm break}$ at 3.26~keV,
which yielded
$\Gamma_{1} = 0.85^{+0.29}_{-0.31}$ and $\Gamma_{2} = 1.47^{+0.44}_{-0.41}$
($\chi^2/\rm d.o.f$ = 16.1/22).

We noted that in \citet{2009AdSpR..43.1036F}, an exponential roll-off component was used
to describe the low-energy part of a spectrum ($<2$ keV) when the spectrum could not
be well fit with a broken PL. We tested the model by adding the roll-off component
({\tt expabs}) to a PL. This model improved the spectral fit over the single PL, but
was worse compared to the models of the extra absorption plus PL  or the broken PL
in all the four cases. This was also true for our fit to the joint {\it NuSTAR} and
{\it Swift}-XRT spectrum (see Table~\ref{joint}).

The {\it Swift} Burst Alert Telescope \citep[BAT;][]{2005SSRv..120..143B}
observed the source with short-exposure observations, where
the hard X-ray flux of the source was below the sensitivity of the BAT
instrument. Therefore no BAT analysis was conducted here.
However the source is included in the {\it Swift}-BAT 70-month hard X-ray
catalogue \citep{2013ApJS..207...19B}. The reported results of
the hard X-ray spectrum (14--195 keV energy range) were a PL with photon index
$\Gamma_{\rm X} = 1.51\pm0.35$
and a flux of
$1.99^{+0.55}_{-0.51}\times 10^{-11}~\rm erg~cm^{-2}~s^{-1}$.

The {\it NuSTAR} spectra in the 3--79 keV energy range were fitted with
an absorbed PL model, where the absorption was fixed at the Galactic value.
The fit yielded photon index $\Gamma_{\rm X} = 1.52\pm0.04$,
an unabsorbed flux (derived using {\tt cflux} model) of
$(3.05\pm0.14)\times10^{-11}~\rm erg~cm^{-2}~s^{-1}$,
with $\chi^2/\rm d.o.f$ = 342.8/390. The photon index obtained with the
absorbed PL model is the same as $\Gamma_{2}$ obtained from the broken PL model
for the joint {\it NuSTAR} and {\it Swift}-XRT spectra in
the $0.3-79$ keV energy range.
We also searched for any hourly variability in the {\it NuSTAR} data,
but no apparent variations were found.

\begin{figure}
\begin{center}
{\includegraphics[width=0.7\linewidth,angle=-90]{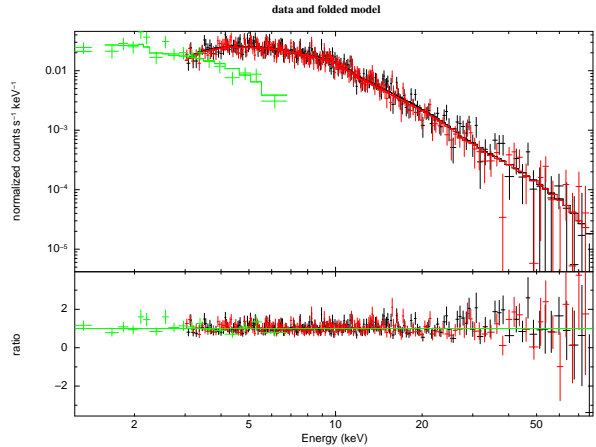}}
\end{center}
\caption{{\it NuSTAR} (red and black points) and {\it Swift}-XRT (green points) spectra and residuals of \4c50, simultaneously fitted with a broken PL.}
\label{spectra}
\end{figure}

The simultaneous observations of \4c50 with {\it NuSTAR} and {\it Swift}-XRT
were performed on 2015 December 14. We thus studied the X-ray spectrum of \4c50
over the wide energy range of $0.3-79$ keV. The broad-band spectrum was
fitted with a PL, a PL plus extra absorption component, a PL plus
exponential roll-off and a broken PL.
In all models, the absorption component ({\tt tbabs}) was fixed at the
Galactic value. The best-fit spectral parameters obtained from
the simultaneous fitting are given in Table \ref{joint}.
The cross-calibration uncertainties between the three telescopes
({\it NuSTAR} FPMA, FPMB, and {\it Swift}-XRT) were considered by
adding a multiplicative constant in the model, which was frozen at 1
for the FPMA spectrum and free to vary for the FPMB and XRT spectra.
The PL model provided an acceptable fit for the joint spectrum
with $\Gamma_{\rm X} = 1.51^{+0.04}_{-0.03}$ and
$\chi^2/\rm d.o.f$ = 373.1/407, while the addition of an extra absorption
at the redshift of the source improved the fit by $\Delta\chi^2 \sim 9$
for the loss of one extra degree of freedom (significance of the extra
component is $\sim 2.9\sigma$).
The extra absorption column density obtained by this fit
is $5.47^{+3.50}_{-3.10} \times 10^{22}~\rm cm^{-2}$.
The PL plus exponential roll-off model for the broad-band spectrum
was marginally as good as the PL plus extra absorption component
($\chi^2$/d.o.f. =366.7/406; see Table~\ref{joint}). The broken PL model further improved the spectral
fit compared to the PL plus the extra absorption model and PL plus exponential roll-off
model. The spectrum and model fit are shown in Figure \ref{spectra}.
The difference in the cross-calibration between FPMA and FPMB was $< 4\%$
in all models, while for the XRT spectrum it was slightly larger but
always less than $13\%$. This difference became larger ($\sim 23\%$)
when a single PL model was used.

\begin{table}
\small
\setlength{\tabcolsep}{12.0pt}
\caption{Simultaneous fit of {\it NuSTAR} and {\it Swift}-XRT data}
\begin{tabular}{@{}lll@{}}
\hline
\hline
Model & Parameter & Value \\
\hline
PL & $\Gamma_{\rm X}$ & $1.51^{+0.04}_{-0.03}$ \\
   & Flux ($0.3-79$ keV) & $3.28^{+0.14}_{-0.13}$ \\
   & $\chi^2/\rm d.o.f$ & 373.1/407 \\
\hline
PL +           & $\Gamma_{\rm X}$ & $1.56\pm0.05$\\
Extra absorber & $\rm N_{\rm H}^{\it z}(\times 10^{22}~\rm cm^{-2})$ & $5.47^{+3.50}_{-3.10}$ \\
               & Flux ($0.3-79$ keV) & $3.21\pm0.14$\\
               & $\chi^2/\rm d.o.f$ & 363.9/406\\
\hline
PL +           & $\Gamma_{\rm X}$ & $1.63^{+0.09}_{-0.08}$\\
Exponential    & $E_{\rm f}$(keV) & $0.87^{+0.60}_{-0.57}$ \\
roll-off       & Flux ($0.3-79$ keV) & $3.09^{+0.18}_{-0.17}$\\
                     & $\chi^2/\rm d.o.f$ & 366.7/406\\
\hline
Broken PL & $\Gamma_{1}$ & $0.05^{+0.88}_{-1.71}$\\
          & $E_{\rm break}$ & $2.13^{+0.62}_{-0.31}$ \\
          & $\Gamma_{2}$ & $1.52^{+0.04}_{-0.03}$ \\
          & Flux ($0.3-79$ keV) & $3.23^{+0.14}_{-0.13}$ \\
          & $\chi^2/\rm d.o.f$ & 360.5/405 \\
\hline
\end{tabular}
\tablecomments{
The errors are at the 90\% confidence level. Flux is in unit of $10^{-11}~\rm erg~cm^{-2}~s^{-1}$ and corrected for the absorption.}
\label{joint}
\end{table}

\section{Discussion}

\subsection{Gamma-ray Properties}
\label{gama_discu}

We have studied the $\gamma$-ray properties of \4c50 by analysing
the {\it Fermi}-LAT data, and confirmed
the prediction in \citet{2010A&A...519A...5A} that \4c50 is
a luminous $\gamma$-ray emitter.
The observed $\gamma$-ray photon index of \4c50\ has a range
of $\Gamma\approx 2.4-3.0$ (see Table~\ref{gamma}),
which is roughly consistent with those of the {\it Fermi}-LAT
$\gamma$-ray FSRQs
($\langle\Gamma\rangle\approx 2.4-2.5$; \citealt{2015ApJ...810...14A}).
From the temporal analysis, we found that \4c50 has been in an
active state since 2013 July. During the active period,
the $\gamma$-ray flux increased by $>3$ times
compared to the quiescence level and the emission was harder.
Moreover, two distinct $\gamma$-ray
flares were well seen in the $0.1-1$ GeV light curve during this period.

Our temporal analysis has shown that \4c50\, exhibited variability on
the time scale of as low as several hours, which is not commonly seen for high
redshift blazars. A blazar jet is produced at the central region
around the super-massive black hole (SMBH), and as
the inner region can not be resolved with current telescopes
(note that thus far,
M87 is the only source resolved with the current observing facilities,
which reaches several Schwarzschild radius, see \citealt{2011Natur.477..185H}),
variability is a useful feature for probing this region.
Given the variability timescale of \4c50, the causality
implies that the size of the emission region is
$R = t_{var} c\delta/(1+z) = 4.2 \times 10^{14} (\delta/10)(t_{var}/1\ {\rm h})\ {\rm cm} = 1.36 \times 10^{15}\ {\rm cm}$ (taking $t_{var}\simeq 4$ h and $\delta=7.9$;
for the $\delta$ value, see below), which is comparable to
the Schwarzschild radius.
The central BH mass of this source is
$\approx4.68\times10^{9}$ M$_{\odot}$ and corresponding Schwarzschild radius
is $1.38\times 10^{15}$ cm \citep{2010A&A...519A...5A}.
Assuming the low-energy X-ray emission is produced from the same region,
the $\gamma$-ray photons could be absorbed by X-ray photons through the
pair production effect. The strength of this absorption is mainly
dependent on the X-ray energy
density, which will decrease if emission is relativistic Doppler beamed.
Therefore, the observed
$\gamma$-ray and X-ray data can be used to constrain the jet Doppler factor. Because
the $\gamma$-ray photons actually escape from the emission region,
the several-hours timescale constrains the lower limit of the beaming factor,
$\delta \ge 11.8 [(1\ {\rm h}/t_{\rm var}) (1\ {\rm keV}/\epsilon_{X}) (L_{\epsilon_{X}}/10^{46}\ {\rm erg~s}^{-1})]^{1/4}$,
where $\epsilon_{X} E_{\gamma}= 20.61 (\delta /10)^{2}$ \citep{1995MNRAS.273..583D}.
Considering $\gamma$-ray photons with energies of $\sim$1~GeV and
the X-ray luminosity of the source obtained in this study,
the Doppler beaming factor $\delta \ge 7.9$ ($t_{var}\sim 4~{\rm h}$).
The VLBI observations show that the apparent superluminal motion reaches $\beta_{app}=6.3c$ and the central jet changes the direction about $\sim 100^{\circ}$ \citep{2007A&A...476L..17A, 2014A&A...566A..26M}. Combining these with the assumption of $\delta=7.9$, we have estimated the viewing angle $\theta=7.9^{\circ}$ and the bulk Lorentz factor $\Gamma_{bulk}=6.5$, which suggest that the jet is highly relativistic and has a small viewing angle with respect to our line of sight.

\subsection{X-ray Properties}
\label{subsec:xray_discu}

We have investigated the X-ray properties of \4c50 using the {\it Swift}-XRT
and {\it NuSTAR} observations. The source showed variability in the
long-term {\it Swift}-XRT light curve.
While its intensity was
at the high end of the variation range during the $\gamma$-ray flaring period
(Figure~\ref{lc}; only the 2014 February {\it Swift} observation was
conducted in the time period), no significant correlated activity was seen.
The X-ray spectral parameters obtained in the active period did not have
drastical changes either.
We considered that
X-rays and $\gamma$-rays are produced from the IC scattering radiation by
the same electron population. Because the cooling timescale of electrons
in the lower energy part (in X-rays) is
longer than the timescale of higher energy part (in $\gamma$-rays),
one can expect that the X-ray variability timescale
would be longer than that of the $\gamma$-rays.

The spectral flattening of the soft X-ray spectrum has been widely found
in high-redshift radio loud quasars
\citep[e.g.][and references therein]{2006MNRAS.368..985Y}. The flattening
may be due to either the intrinsic absorption with column densities of
the order of $10^{22} -10^{23}~\rm cm^{-2}$ or the low energy cut-off
in the energy distribution of electron population in the
jet \citep{2001MNRAS.323..373F,2001MNRAS.324..628F,2004MNRAS.350L..67W,2004MNRAS.350..207W}.
In the excess absorption scenario,
high $\rm N_{\rm H}^{\it z}$ may be the dense plasma in
form of a wind or outflow \citep{1999MNRAS.308L..39F}.
However in the radio-loud quasars like \4c50, the relativistic jet along
the line-of-sight can remove the gas column efficiently. Indeed, the VLBI
observations \citep{2007A&A...476L..17A}
have revealed a jet toward the Earth, suggesting that the excess absorption
scenario is not likely the case.

If there is a low energy cut-off in the energy distribution of the electron
population, a spectrum is expected to flatten in the soft energy
band \citep{2001MNRAS.323..373F,2007ApJ...665..980T,2007ApJ...669..884S}.
This scenario requests a broken PL model, where the cut-off in the
soft X-ray band can be naturally explained as the intrinsic curvature of
the spectrum near the low-energy end of the IC component.
The soft X-ray flattening is then an intrinsic feature of a source.
Among the four examined observations of \4c50, the spectra were relatively
well described by the broken PL model of $\Gamma_{1}\sim -0.3$ -- +0.9 below
the break energy $E_{\rm break} =$ 2.1 -- 3.3 keV, and
$\Gamma_{2} \sim$ 1.4 -- 1.8.
The simultaneous observations of \4c50 by {\it Swift} and {\it NuSTAR}
showed that the broad-band X-ray spectrum is better modelled by a broken PL
than by a PL or a PL plus extra absorption model.
We found $\Gamma_{1} = 0.05^{+0.88}_{-1.71}$ below the break energy,
$E_{\rm break} = 2.13^{+0.62}_{-0.31}$ keV, and $\Gamma_{2} = 1.52^{+0.04}_{-0.03}$.
In the 14--195 keV energy range of the {\it Swift}-BAT, the source
was found to have photon index $\Gamma_{\rm X} = 1.51\pm0.35$
\citep{2013ApJS..207...19B},
which is well in agreement with $\Gamma_{2}$ obtained in our broad-band fit.
Thus we suspect that the flattening is likely the intrinsic feature of
the source.
This possibility is supported by the broad-band SED modelling
(see the following Section 5.3 and Table~\ref{tab:modfit}).
From the modeling, it can be known that the low energy of
non-thermal electrons is about $\gamma_{min}\sim1.1$ and the Doppler beaming
factor $\delta\sim22.5$ for the active state. The electrons around
the minimum energy will IC scatter external seed photons, and emit
at $\nu_{IC}\approx(4/3)\delta\Gamma_{jet}\gamma_{min}^{2}\nu_{ext}/(1+z)\sim2.6$ keV (assuming $\Gamma_{jet}=\delta$),
which is roughly consistent with observations
(similarly, we have $\nu_{IC}\approx2.1$ keV for the quiescent state).

\subsection{Spectral Energy Distribution Fitting}
\label{sed_discu}

\begin{figure*}
\begin{center}
{\includegraphics[width=0.8\linewidth]{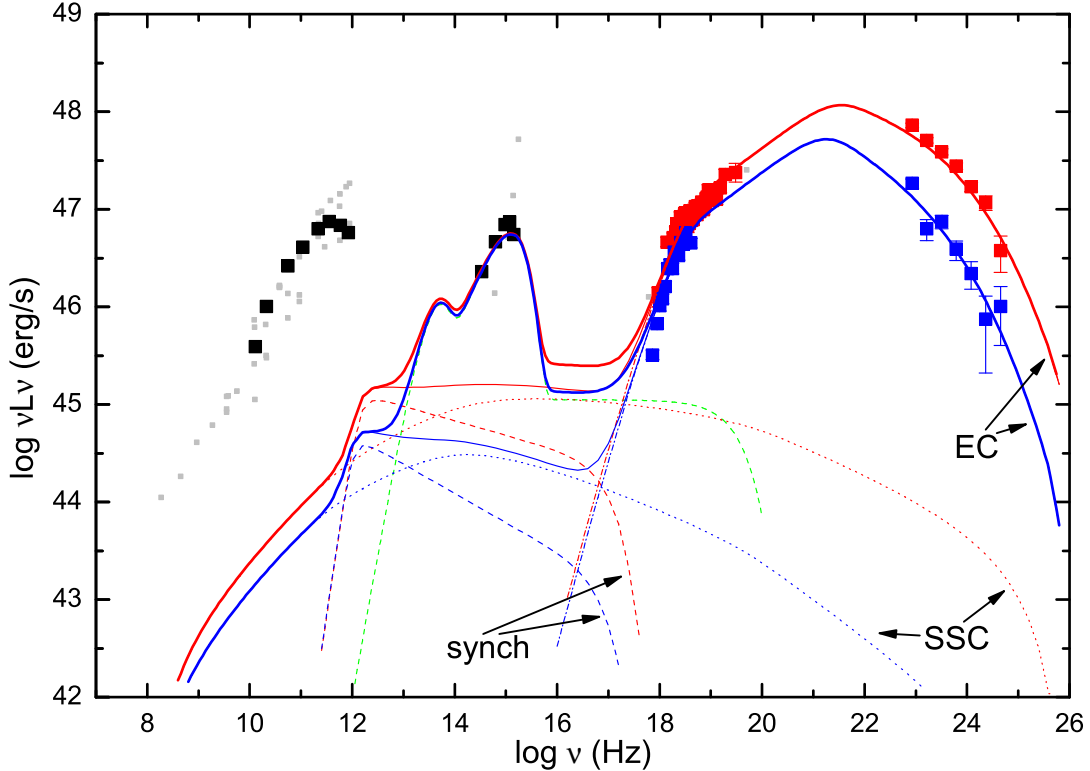}}
\end{center}
\caption{Broad-band SED of \4c50, the black and grey data points
are separately from \citet{2010A&A...519A...5A} and NED.
The red points (lines) stand for active states, while the blue
ones for quiescent states. Solid lines are for EC/BLR model.
The central accretion disk/corona/dust torus emission component is
represented using green dash line.
The red/blue dash, dot and dash dot lines are synchrotron, SSC and EC
emission in active/quiescent states, respectively.}
\label{sed}
\end{figure*}

We collected the archival radio and optical data for \4c50
from \citet{2010A&A...519A...5A} and
NED\footnote{\url{http://ned.ipac.caltech.edu/}} respectively. These data were
combined with X-ray and $\gamma$-ray data in this work and
the broad-band SED of the source is shown in Figure~\ref{sed}.
In this SED, emission from the relativistic jet dominated except at
optical wavelengths.
The optical emission reached a peak luminosity of $\sim10^{47}$ erg s$^{-1}$
and appeared as a significant bump, which should be thermal arising from
the optically thick accretion disk. As \4c50 hosts a very massive BH,
the thermal disk emission reaches $\sim30\%$ of the Eddington
limit \citep{2010A&A...519A...5A}. The broad-band SED is not simultaneous,
except the two sets of {\it Swift} X-ray and
the corresponding \fermi\ $\gamma$-ray data in the active and
quiescent states (the red and blue squares, respectively, in Figure~\ref{sed}).
Nevertheless, we modelled the broad-band SED by using a standard blazar
emission model: one zone synchrotron plus inverse Comptonization model.
This model was widely used in blazar SED
modelling \citep[e.g.,][]{2010MNRAS.402..497G, 2012ApJ...748..119C,2017ApJ...842..129C}.
The emission region is assumed to be a homogeneous sphere with radius $R$
embedded in the magnetic field $B$. A broken power-law electron energy
distribution,

\begin{table*}
\centering
\setlength{\tabcolsep}{4.0pt}
\caption{SED Model parameters for 4C 50.11}
\begin{tabular}{@{}lcccccccccccccc@{}}
\hline
\hline
State & $B$ & $R$ & $N_0$ & $\delta$ & $\gamma_0$ & $\gamma_{min}$ & $p_1$ & $p_2$ & $P_{jet}^{e-p}$ & $P_{jet}^{e-e^{+}}$ & $L_{diks}$\\
      & Gs & $10^{15}$cm       &        &          &            &               &                 &      &  $10^{48}$erg/s & $10^{45}$erg/s  &  $10^{47}$erg/s      \\
\hline
   high &    0.8 &    3.80 &    0.628E+07 &   22.16 &   34.80 &    1.1 &   2.2 &   3.3 &   4.616 &   5.930 &   1.8 \\
   low &    0.8 &    4.29 &    0.422E+07 &   20.00 &   30.62 &    1.1 &   2.2 &   3.6 &   3.203 &   3.973 &   1.8 \\
\hline
\end{tabular}
\label{tab:modfit}
\end{table*}

\begin{equation}
N(\gamma )=\left\{ \begin{array}{ll}
                    N_{0}\gamma ^{-p_1}  &  \mbox{ $\gamma_{\rm min}\leq \gamma \leq \gamma_{0}$} \\
            N_{0}\gamma _{\rm 0}^{p_2-p_1} \gamma ^{-p_2}  &  \mbox{ $\gamma _{\rm 0}<\gamma\leq\gamma_{\rm max}$,}
           \end{array}
       \right.
\label{Ngamma}
\end{equation}
was assumed in our calculation. The parameters of this model include
the radius $R$ of the blob, the magnetic field strength $B$, electron
break energy $\gamma_{0}$, the minimum and maximum energy, $\gamma_{\rm min}$
and $\gamma_{\rm max}$, of the electrons, the normalization of
the particle number density $N_{0}$, and the indices $p_{1,2}$ of
the broken power law particle distribution, the jet Doppler
factor (assumed to be equal to the bulk Lorentz factor), and
the spectrum of the external seed photons. The frequency and luminosity can
be transformed from the jet frame to observational frame as:
$\nu=\delta\nu'/(1+z)$ and $\nu L_{\nu}=\delta^{4}\nu'L_{\nu'}'$, where
the Doppler factor $\delta=1/\left[\Gamma\left(1-\beta\cos\theta\right)\right]$,
and the prime represents the value measured in the jet frame. The
synchrotron self-absorption and the Klein-Nishina effect in the
IC scattering were properly considered in our calculations.
Both the self-synchrotron Compton (SSC) scattering and external Compton (EC)
scattering (external seed photons from the BLR and dusty torus were taken into
account) were included in the calculation of the Compton scattering in
the blob.

As mentioned above, the optical emission is multi-temperature
annular blackbody radiation arising from the accretion disk, which was
modelled with a standard optically thick, geometrically thin disk
(Shakura \& Sunyaev 1973). Above the accretion disk, the corona reprocessed
a fraction of disk luminosity (fixed at a level of 10\%)
and had a power-law spectrum with cut-off energy 150 keV (we fixed the spectral
index $\alpha=1.0$). Because the accretion disk's
radiation is de-beamed in the jet comoving frame, seed photons from it were
not important and thus not included in the EC scattering.
In our SED modelling, the luminosities of the BLR and dust torus were
assumed to be a fraction of the disk luminosity, 10\% and 50\%,
respectively \citep{2008MNRAS.387.1669G}. The radii of the BLR and torus were
$R_{BLR}=10^{17}L_{disk,45}^{1/2}=0.43$ pc and
$R_{torus}=2.5\times10^{18}L_{disk,45}^{1/2}=10.9$ pc, respectively
($L_{disk,45}=179.0$ is the disk luminosity in units of $10^{45}$ erg s$^{-1}$;
see \citealt{2010A&A...519A...5A}). In this case, the external photon energy
densities are typical values $U_{BLR}=2.65\times10^{-2}$ erg cm$^{-3}$
and $U_{torus}=2.12\times10^{-4}$ erg cm$^{-3}$. The size of the emitting
region was assumed to be equal to the radius of a circular conic section,
$R=\psi R_{diss}$ ($R_{diss}$ is the distance of the emission region
from the central black hole, where $\psi=0.1$;
see \citealt{2008MNRAS.387.1669G}). The variability timescale can be used to
set an upper limit on the emission size due to the causality,
$R\lessapprox c\Delta t\delta/(1+z)$. During our SED modeling,
the minimum variability timescale ($\Delta t\approx$4h) was used for
estimating the size of the emission region for the active state. Note that
the Doppler factor estimated in Section~\ref{gama_discu}, $\delta\gtrsim7.9$,
was the lower limit to avoid the absorption of $\gamma$-ray photons
through electron pair production effect.

In Figure~\ref{sed}, we show the model fits to the SEDs in both
the active and quiescent states, with seed photons dominantly coming from
the BLR.  The model parameters are given in Table~\ref{tab:modfit}.
From the jet bolometric luminosity $L_{jet}$, we can obtain
the jet non-thermal radiation power \citep{2014Natur.515..376G},
$P_{\rm rad}\approx2L_{\rm tot}/\delta^{2}=3.1\times10^{46}$ erg s$^{-1}$
for the active state, which is about $\sim 17\%$ of the disk luminosity
of $1.8\times10^{47}$ erg s$^{-1}$. The jet radiative efficiency is
believed to be on order of $P_{\rm rad}/P_{\rm jet}\sim$10\%, which holds
for AGNs, gamma-ray bursts, and even for black hole X-ray
binaries \citep{2012Sci...338.1445N, 2013ApJ...774L...5Z, 2014ApJ...780L..14M},
which gives a jet power,
$P_{\rm jet}\thickapprox10P_{\rm rad}=3.1\times10^{47}$ erg s$^{-1}$,
larger than the disk luminosity. This suggests that the jet launching
processes and the way of transporting energy from vicinity of the black hole
must be very efficient. Actually, having the model parameters, the jet power
can be calculated as,
$P_{\rm jet}\simeq\pi R^{2}\beta \Gamma^{2}cU_{\rm tot}'$, where
the total energy density measured in the rest frame of the jet,
$U_{\rm tot}^\prime=U_{\rm e}^\prime+U_{\rm B}^\prime+U_{\rm p}^\prime$.
The energy density for electrons $U_{\rm e}^\prime=m_{\rm e}c^{2}\int N(\gamma)\gamma d\gamma$, while the proton energy density
$U_{\rm p}^\prime=U_{\rm e}^\prime(m_{\rm p}/m_{\rm e})/\langle\gamma\rangle$
if charge neutrality for pure hydrogen
plasma is assumed. The estimated values for the jet powers are given in
Table~\ref{tab:modfit}. It can be seen that the jet power $P_{jet}$ is
larger than the disk luminosity $L_{disk}$ by more than one magnitude
and even larger than that of accretion power
$P_{acc}=L_{disk}/\eta\approx(0.6-1.8)\times10^{48}$ erg s$^{-1}$,
where the radiative efficient of the accretion disk is assumed to
be $\eta\approx0.1-0.3$ \citep{2014Natur.515..376G}.
However, we note that the minimum electron energy in our fitting is small, 
$\gamma_{min} = 1.1$ (Table~\ref{tab:modfit}). The small value of $\gamma_{min}$
may result in the overestimation of the jet power. In SED modeling, the reproduction of X-ray emission
is important to constrain $\gamma_{min}$. In \citet{2014ApJ...788..104Z,2015ApJ...807...51Z}, X-ray is produced
through SSC mechanism, in which $\gamma_{min}$ is much larger than the unit. In our SED modeling, we failed to
model the X-ray with the SSC emission; instead, following \citet{1998MNRAS.301..451G,2010MNRAS.402..497G}, we modeled
the X-ray through EC emission and therefore obtained a small $\gamma_{min}$ (similar to \citealt{1998MNRAS.301..451G,2010MNRAS.402..497G}).
In addition, it should be noted that the estimated jet power is largely dependent
on the assumed jet components. For exmaple, if the jet is mainly composed
of electron-positron pairs instead of electron-proton plasma,
the jet powers will be significantly decreased and smaller than
the accretion disk luminosity for both the active and quiescent states
(see Table \ref{tab:modfit}).


\section{Summary}

We have studied the $\gamma$-ray and X-ray properties of the high redshift blazar \4c50
from analysing the {\it Fermi}-LAT, {\it Swift}, and {\it NuSTAR} data.
The main results are summarized as the following.

\begin{itemize}

\item From {\it Fermi}-LAT monitoring, the source was found to be in
an active state since approximately MJD 56482. During the state,
the source's $\gamma$-ray flux increased as largely as nearly
one order of magnitude (averaged on 5 days bin) compared to the quiescent
level. In addition, the $\gamma$-ray spectra appeared harder
during the active period.
We also found that the $\gamma$-ray variability can be resolved
on several hours level.  The property has helped
constrain the physical properties of the jet of the blazar.

\item The source showed flux variability in the {\it Swift} and {\it NuSTAR}
data we have analysed, but no obvious flux enhancement or spectral changes
related to the $\gamma$-ray active state were seen. As long as
the data quality allows, we have found that a broken power law
provided the best fit to the broad-band X-ray spectra, with an extremely flat
spectrum ($\Gamma \sim 0.1$) below the break energy,
$E_{\rm break} \sim 2.1~{\rm keV}$,
and a flat spectrum above the break energy. This spectral feature is likely
due to the low-energy cutoff in the energy distribution of the photon-emitting
electron population.

\item We have constructed the broad-band SED for \4c50,
though not simultaneous, and provided a
model fit (one-zone synchrotron plus inverse Comptonization model)
to the SED.  From modelling, properties of the emission region were derived.

\end{itemize}

\acknowledgments

We acknowledge the use of data from \fermi~Science Support Center (FSSC), and
{\it Swift}, {\it NuSTAR} data from the High Energy Astrophysics Science
Archive Research Center (HEASARC), at NASA’s Goddard Space Flight Center.
This research
has made use of the High Performance Computing Resource in the Core Facility for
Advanced Research Computing at Shanghai Astronomical Observatory,
the {\it NuSTAR} Data Analysis Software (NUSTARDAS) jointly
developed by the ASI Science Data Center (ASDC, Italy) and the California
Institute of Technology (Caltech, USA).

This research was supported by the National Program on Key Research
and Development Project (Grant No. 2016YFA0400804),
the National Natural Science Foundation of China for Youth (11603059),
the National Natural Science Foundation of China (11373055, 11633007, 11233006, U1431123),
and the CAS grant (QYZDJ-SSW- SYS023).
VJ acknowledges the financial support
from Chinese Academy of Sciences through President's International Fellowship
Initiative (CAS PIFI, Grant No. 2015PM059).
Z.W. acknowledges the support by the CAS/SAFEA International Partnership
Program for Creative Research Teams.



\bibliography{ms-4c50_V2}

%
%



\end{document}